\documentclass[prl,twocolumn,showpacs,amsmath,amssymb,superscriptaddress]{revtex4-1}
\usepackage[english]{babel}
\usepackage{amsmath}
\usepackage{graphicx, nicefrac}
\usepackage{float}
\usepackage{siunitx}

\usepackage{dcolumn}
\usepackage{bm,graphicx}
\usepackage{etoolbox}

\usepackage[colorlinks]{hyperref}
\usepackage{url}
\usepackage[colorlinks]{hyperref}
\usepackage[normalem]{ulem}
\usepackage[dvipsnames,usenames]{xcolor}
\usepackage[version=3]{mhchem} 

\newcommand{\bns}{\ce{BaNiS2}}

\newcommand{\K}[1]{\SI{#1}{\kelvin}}
\newcommand{\mev}[1]{\SI{#1}{\milli\electronvolt}}
\newcommand{\ev}[1]{\SI{#1}{\electronvolt}}

\newcommand{\GM}{$\Gamma$--M}
\newcommand{\GX}{$\Gamma$--X}

\begin{document}

\title{Optical conductivity signatures of open Dirac nodal lines}

\author{David Santos-Cottin}
\email[]{david.santos@unifr.ch} 
\affiliation{LPEM, ESPCI Paris; CNRS; PSL University, 10 rue Vauquelin, F-75005 Paris, France}
\affiliation{Sorbonne Universit\'e, CNRS, LPEM, F-75005 Paris, France}
\affiliation{Department of Physics, University of Fribourg, 1700 Fribourg, Switzerland}

\author{Michele Casula}
\affiliation{IMPMC-Sorbonne Universit\'e, CNRS and MNHN, 4, place Jussieu, 75005 Paris, France}

\author{Luca de' Medici}
\affiliation{LPEM, ESPCI Paris; CNRS; PSL University, 10 rue Vauquelin, F-75005 Paris, France}
\affiliation{Sorbonne Universit\'e, CNRS, LPEM, F-75005 Paris, France}

\author{F. Le Mardel\'e}
\affiliation{Department of Physics, University of Fribourg, 1700 Fribourg, Switzerland}

\author{J.~Wyzula}
\affiliation{LNCMI-EMFL, CNRS UPR3228, Univ. Grenoble Alpes, Univ. Toulouse, Univ. Toulouse 3, INSA-T,  Grenoble and Toulouse, France}

\author{M.~Orlita}
\affiliation{LNCMI-EMFL, CNRS UPR3228, Univ. Grenoble Alpes, Univ. Toulouse, Univ. Toulouse 3, INSA-T,  Grenoble and Toulouse, France}
\affiliation{Institute of Physics, Charles University in Prague, CZ-12116 Prague, Czech Republic}

\author{Yannick Klein}
\affiliation{IMPMC-Sorbonne Universit\'e, CNRS and MNHN, 4, place Jussieu, 75005 Paris, France}

\author{Andrea Gauzzi}
\affiliation{IMPMC-Sorbonne Universit\'e, CNRS and MNHN, 4, place Jussieu, 75005 Paris, France}

\author{Ana Akrap}
\affiliation{Department of Physics, University of Fribourg, 1700 Fribourg, Switzerland}

\author{R. P. S. M. Lobo}
\email[]{lobo@espci.fr} 
\affiliation{LPEM, ESPCI Paris; CNRS; PSL University, 10 rue Vauquelin, F-75005 Paris, France}
\affiliation{Sorbonne Universit\'e, CNRS, LPEM, F-75005 Paris, France}

\date{\today}
\begin{abstract}
We investigate the optical conductivity and far-infrared magneto-optical response of \bns, a simple square-lattice semimetal characterized by Dirac nodal lines that disperse exclusively along the out-of-plane direction. 
With the magnetic field aligned along the nodal line the in-plane Landau level spectra show a nearly $\sqrt{B}$ behavior, the hallmark of a conical-band dispersion with a small spin-orbit coupling gap. 
The optical conductivity exhibits an unusual temperature-independent isosbestic line, ending at a Van Hove singularity.
First-principles calculations unambiguously assign the isosbestic line to transitions across Dirac nodal states.
Our work suggests a universal topology of the electronic structure of Dirac nodal lines. 
\end{abstract}
\pacs{}
\maketitle

%
%
Certain Dirac semimetals belonging to a class of nonsymmorphic square-lattice system~\cite{Schoop2016,Chen2017,Yang2018,Takane2016} outshine the numerous topological systems known today. 
They have Dirac points located on in-plane symmetry lines, in contrast to  high-symmetry points of the Brillouin zone as in graphene.
This  is a promising property for novel concepts of electronic devices, as it makes the topological properties of two-dimensional (2D) materials tunable.
Indeed, it has been shown that electronic correlations effectively modify the dispersion of the Dirac nodes \cite{Li2008,Shao2020,Nilforoushan2020,Gatti2020}. This dispersion typically follows either an open line or a close loop, thus reducing the effective dimension from three to two \cite{Carbotte2016}. 
The challenge is that the most studied Dirac nodal line (DNL) materials have intricate Fermi surfaces with Dirac lines dispersing in different reciprocal space directions.
It is therefore important to identify model systems where the physics of DNLs is as simple as possible. 

\begin{figure*}[!th]
	\includegraphics[width=0.95\linewidth]{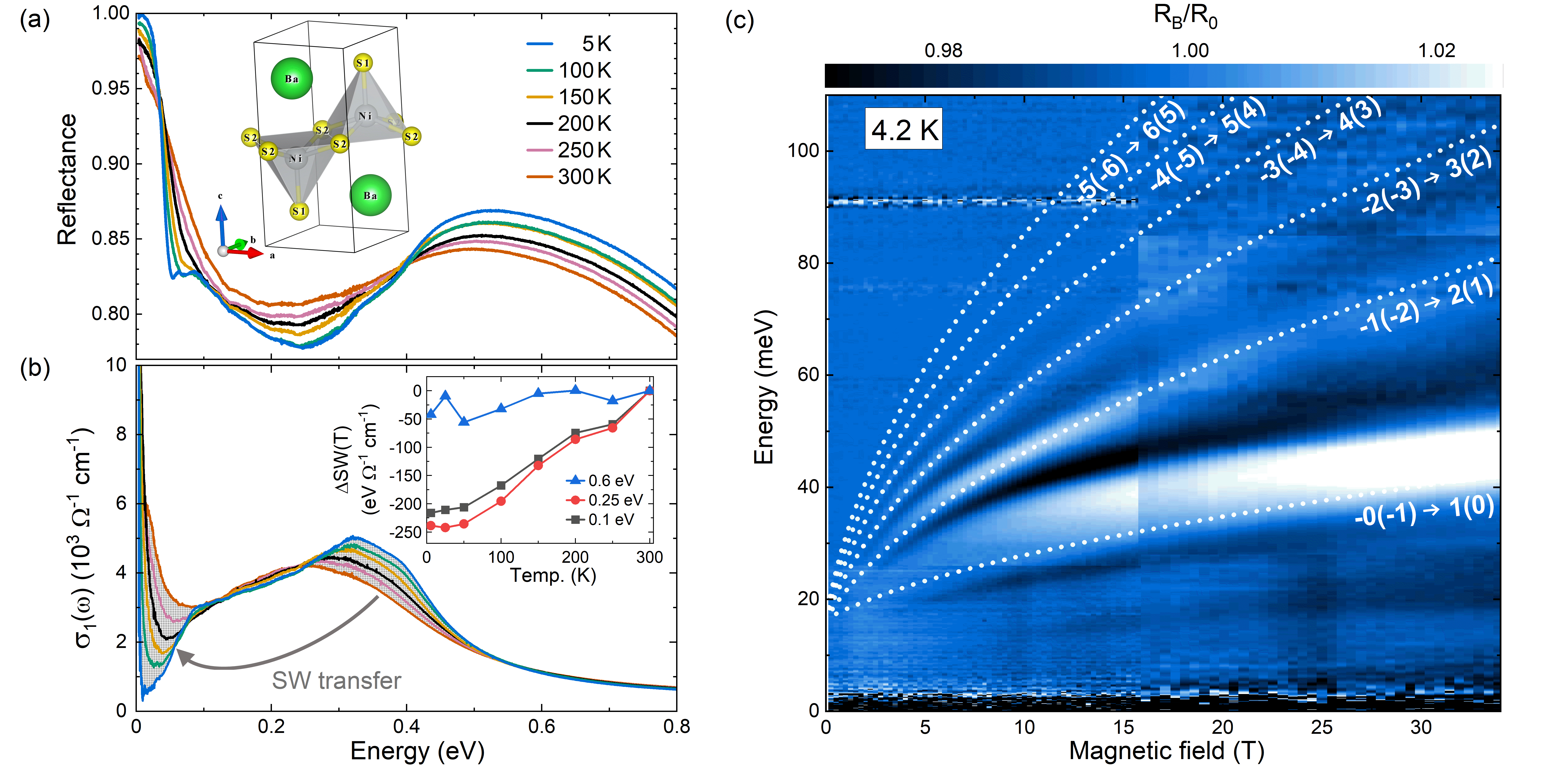}
	\caption{(a) Temperature dependent in-plane reflectivity of \bns\ below 0.8~eV. Inset: Crystal structure of \bns. 
	(b) Temperature dependence of the in-plane real part of the optical conductivity. The grey areas show the SW transfer from high to low energy between 300~K and 5~K.
	The Inset shows the differential spectral weight $\Delta SW(T) = SW(T) - SW_{300K}$, for different upper cut-off energies, as a function of temperature.
	(c) Color-plot of the relative magneto-reflectance, $R_B/R_0$. Lighter areas correspond to
inter-LL excitations, which were assigned to transitions: $-n \rightarrow n+1$ and $-n-1 \rightarrow n$ and fitted using the LL spectrum in Eq. \ref{eq0}.
}
	\label{fig1}
\end{figure*}

\begin{figure*}[!ht]
	\includegraphics[width=0.95\linewidth]{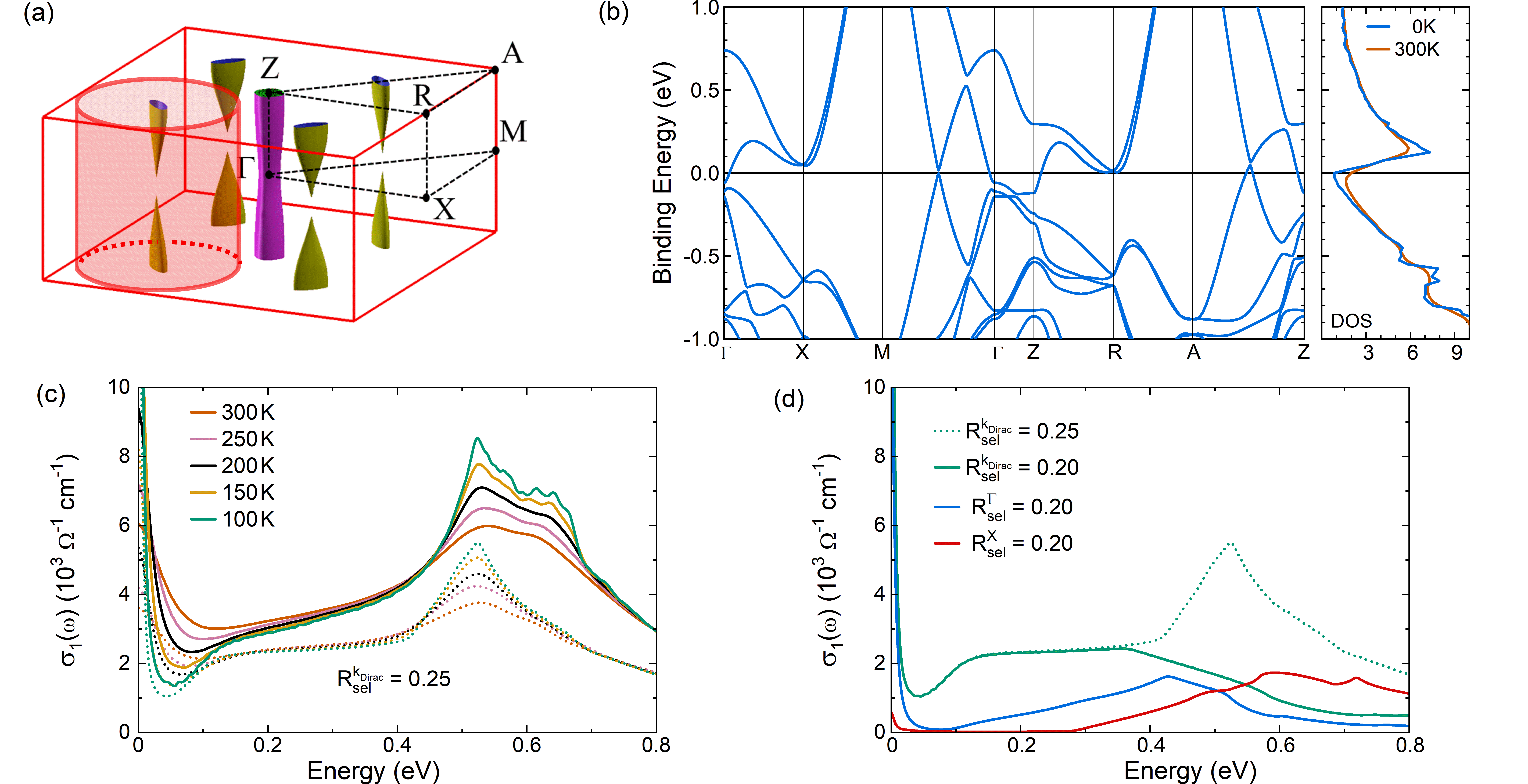}
  	\caption{(a) First Brillouin zone and the HSE-calculated Fermi surface. Hole and
          electron pockets are indicated in yellow and purple,
          respectively. The red cylinder 
          represents the constrained momentum space
          used to isolate the contribution of the nodal line. 
          (b) HSE band structure, and DOS at zero and 300~K.
          (c) Low-energy optical conductivity calculated at different temperatures for the full IBZ (solid 
          lines). Dashed lines show $\sigma_1(\omega)$ for transitions in the 
          restricted $k$-space, defined by the shaded volume in Fig.~\ref{fig2}(a). 
          (d) Full optical conductivity at $100$~K decomposed into 
          different $k$-space contributions, for various cylinders in IBZ. 
          Green, blue, and red lines shows $\sigma_1(\omega)$ for a cylinder centered at $k_\textrm{Dirac}$, $\Gamma$, and X, respectively.
          The radius R$_\textrm{sel}$ of each cylinder is given in crystal units.
	}
  	\label{fig2}
\end{figure*}

\begin{figure*}[!ht]
	\includegraphics[width=0.95\linewidth]{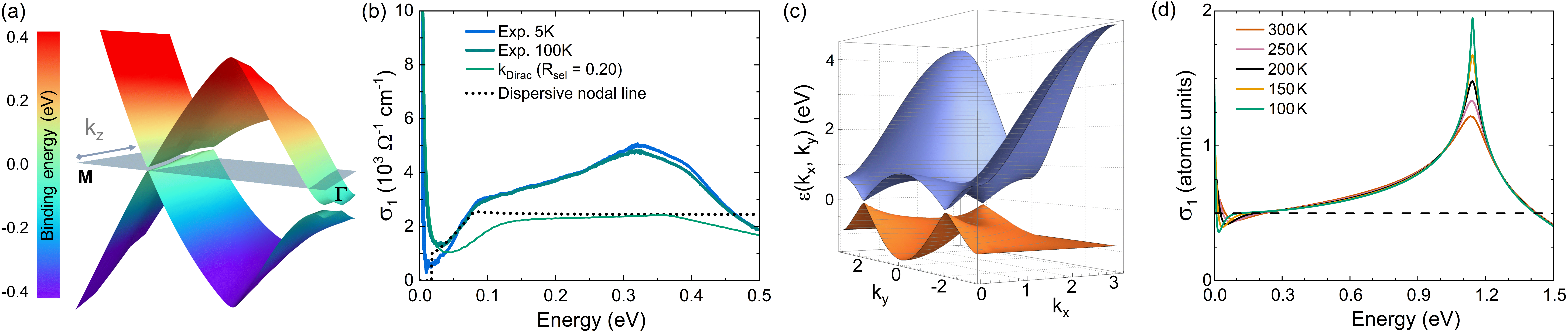}
	\caption{
         (a) 3D view of the connection between the two DNLs dispersing along $k_z$ obtained from HSE calculations. 
	(b) Comparison between the measured  $\sigma_1(\omega)$ at
        \K{10} and \K{100} (thick lines), and the analytic form of
        Eq.~\ref{eq3} (dashed line) with an electric field
        perpendicular to the nodal line dispersion $k_z$. The thin
        green line shows the $k$-space restricted
        theoretical $\sigma_1(\omega)$ at 100~K.
	(c) Dispersion of a 2D model on a square lattice.
	(d) $\sigma_1(\omega)$ of the model in panel (c) at
          different temperatures. The peak corresponds to interband transitions
          between two saddle points (VHSs) of the
          model. The black dashed line is the universal
          value of $\sigma_1(\omega)$ for a perfect conical model with
          4 valleys and 1/2-spin fermions.
     }
	\label{fig3}
\end{figure*}

A key finding in the physics of these materials happened long before any topological effects were known, when Hoffmann and Tremel \cite{Tremel1987} showed that the nonsymmorphic tetragonal $P4/nmm$ space group can lead to dispersive DNLs.
Only a handful of such compounds are actively explored \cite{Takane2016,Ebad2019}. 
Among the less explored compounds, we focus on \bns\ \cite{Grey1970} whose DNLs have been discovered and studied only recently~\cite{Santos-Cottin2016,Santos-Cottin2016a,Nilforoushan2019,Nilforoushan2020}. 
\bns\ is formed by square-lattice layers of puckered NiS$_5$ pyramids shown in the inset of Fig.~\ref{fig1}(a). 
Remarkable features of the band structure are fourfold-symmetry DNLs within the $k_x-k_y$ plane, with Dirac nodes pinned at the Fermi energy at $k_z=0$. 
These nodes slightly disperse along $k_z$ and raise above the Fermi level as $k_z$ increases
(Fig.~\ref{fig3}(a)). 
The resulting nodal lines are open, wrapping the first Brillouin zone (IBZ) torus in the $k_z$ direction. 
Their in-plane momentum position is tunable along the \GM\ path (Fig.~\ref{fig2}(a)) through chemical substitution~\cite{Nilforoushan2020}, and their Fermi velocities can be modulated by external perturbations, such as light pulses~\cite{Nilforoushan2019}.

This ensemble of favorable properties makes \bns\ a model system for probing open DNLs through dynamic charge transport.
In this Letter we report its first magneto-optical conductivity study combined with first-principles calculations.
We find that the nodal lines' contribution to the spectrum has universal optical conductivity fingerprints:
the spectrum has a linear energy dispersion; and
it is temperature ($T$-) independent over an ``isosbestic'' range of frequencies, 
ending at high energy peaks, which do have a $T$-dependence.
The linear dispersion and $T$-independent conductivity is similar to that found in graphene, confirming the effective two-dimensionality of the conical dispersion in \bns~\cite{Falkovsky2007,Falkovsky2007B,Kuzmenko2008,Mak2008,Mak2011,Schilling2017}. 
The $T$-dependent peaks cannot be dissociated from the lower-energy isosbestic range, because the bands generating the Dirac cones are continuous and IBZ-periodic. 
According to the Morse theorem~\cite{Morse1925}, these bands must show saddle points, yielding Van Hove singularities (VHSs) with a logarithmic divergence in the 2D density of states~\cite{VanHove1953}, and leading to $T$-dependent peaks in the optical conductivity once finite lifetime and many-body effects are taken into account~\cite{Mak2011,Mak2014}.
Considering that the above connection between Dirac bands and VHSs is a topological property of a quasi-2D electronic structure, it follows that the features observed in \bns\ are universal fingerprints of open DNL semimetals, expected in layered beyond-graphene materials. 

%
%
 
We measured the in-plane reflectance of 0.5-1 mm-sized \bns\ single crystal platelets \cite{Santos-Cottin2016,Shamoto1995} from 3 meV to 5 eV. We also measured their far-infrared response in fields up to 34~T \cite{SM}. 
First-principles calculations were carried out within  density functional theory in the \textsc{Quantum Espresso} plane-wave implementation \cite{QE-2009,giannozzi2017}. 
We employed a modified hybrid Heyd-Scuseria-Ernzerhof (HSE) functional~\cite{heyd2003hybrid,ge2006erratum,krukau2006influence}, able to accurately reproduce the fermiology of \bns.
From the band structure, we computed the optical conductivity in the Kubo formalism, by explicitly including thermal scattering effects via a Fermi-liquid self-energy $\Sigma=i \Gamma = i \pi (k_B T)^2/(k_B T_0)$, with $T_0=1000$~K \cite{SM,mostofi2014,kubo1957,pruschke1993,tomczak2009}.

%
%

To access the complex optical conductivity, we measured the frequency-dependent reflectance, shown in Fig.~\ref{fig1}(a) \cite{Homes1993,SM}. We calculated the optical conductivity from the reflectance employing Kramers-Kronig relations \cite{Tanner2015}. 
Its real part $\sigma_1(\omega)$, shown in Fig.~\ref{fig1}(b),
contains a highly unusual isosbestic range where it remains $T$-independent from 80 to 250~meV. This isosbestic line is in contrast with the isosbestic point commonly observed in gapped systems.
Such an isosbestic line means that the spectral weight (SW), defined as the area underneath $\sigma_1 (\omega)$, depleted from low frequencies at low temperatures gets transferred to high energies, skipping over the 80 to 250~meV range. 
The inset of Fig.~\ref{fig1}(b) illustrates that the SW lost from dc to about 100~meV is the same as the one lost from dc to about 250~meV. Only then does it start to get compensated until it is fully recovered by 600~meV.

Focusing on the 5~K curve, we see that $\sigma_1(\omega)$ contains several contributions. 
A very narrow Drude component is followed by a roughly linear increase up to 100~meV. 
At this point, another linear dispersion begins with a smaller slope and a finite zero-energy intercept. 
This second linear dispersion is the isosbestic line where $\sigma_1(\omega)$ at all temperatures overlap.
It terminates in a pair of broad, partly overlapping peaks, at 0.3 and 0.4~eV.

Applying the magnetic field along the nodal-lines direction (001), a series of excitations is observed in the relative magneto-reflectivity spectra, $R_B/R_0$, Fig.~\ref{fig1}(c) 
\footnote{The reason we plot the reflectivity maxima is because $\epsilon_2$ is larger than $\epsilon_1$, and therefore dominates the reflectivity response.} \cite{SM}.
The maxima correspond to the transition energies and are electric-dipole excitations, with selection rules $\Delta |n| = \pm 1$, between pairs of Landau levels (LLs) of 2D massive Dirac electrons:
\begin{equation}
\varepsilon_{\pm n}(B) =\pm\sqrt{2 \hbar e v_{1}v_{2} B n + \Delta^{2}}, \;\; n =0,1,2\ldots.
\label{eq0}
\end{equation}
The most pronounced line corresponds to the lowest interband excitations $-1\rightarrow 0$  and $0\rightarrow1$. 
The latter dominates in the hole doped parts of the nodal lines (away from $k_z=0$).   
At lower energies, intraband (cyclotron-resonance-like) inter-LL excitations appear. 
The extracted gap is $2\Delta = 16$~meV, and the asymptotic velocity parameter, $v=\sqrt{v_1 v_2} = 1.55\times 10^5$~m/s
, where $v_1$ and $v_2$ are velocities for the two in-plane directions. 
\bns\ is the nodal line system having the smallest gap, 3 times smaller than the one in ZrSiSe~\cite{Shao2020}, twice smaller than that of ZrSiS \cite{Schoop2016}, and 7 times smaller than the gap in NbAs$_2$ \cite{Shao2019}. 
The velocity parameter $v$ is 1.6 times lower than in ZrSiSe~\cite{Shao2020}.

The optical conductivity is dominated by interband contributions, and these can be best understood by knowing the exact band structure. 
Figures~\ref{fig2}(a) and \ref{fig2}(b) display the calculated Fermi surface, band structure and the density of states (DOS) of \bns\ around the Fermi level. 
The computed band structure yields two pockets at the Fermi energy ($E_F$), in agreement with previous results \citep{Santos-Cottin2016, Klein2018}. 
Halfway through the \GM\ line, there are two spin-degenerate linear bands. 
These bands cross at $E_F$ and $k_\textrm{Dirac}$ at $k_z=0$, forming a well-defined Dirac cone, a common signature of the square-net family \cite{Tremel1987, Yang2018}. 
Because of the SOC, the Dirac cone is slightly gapped, 2$\Delta_{calc} =$ \mev{18}, in 
agreement with magneto-optics.
The Dirac point progressively shifts to higher energies as $k_z$ increases, leading to a hole-like Fermi pocket along the Z--A line, compensated by an electron pocket centered at $\Gamma$.
In contrast to other square net lattice members, \bns\ shows no additional low-energy Dirac structures in the \GX\ or Z--R directions at $E_F$.
This makes \bns\ a cleaner case, and a good candidate to isolate the signature of the dispersive nodal line.

Full lines in Fig.~\ref{fig2}(c) display the calculated optical conductivity.
For $T < \K{200}$, the Drude width is small enough to reveal a minimum around \mev{80}, followed by a linear increase of $\sigma_1(\omega)$ up to $\sim$ \mev{120}.
In the 200--\mev{450} range, $\sigma_1(\omega)$ collapses into a linear-in-energy response, roughly independent of temperature.  
Two peaks appear around 550 and \mev{650}. These features are enhanced at lower temperatures, and they appear at  energies $\sim 1.5$ times higher than those experimentally determined.
This discrepancy points to sizable band renormalization effects, driven by electronic correlations, which are not perfectly captured by the modified HSE functional, and to exitonic many-body effects beyond single-particle description~\cite{Mak2011,Mak2014}. 
Nevertheless, all the important characteristics of the data are reproduced by the first-principles results: 
(i) a Drude peak and a small gap at low temperatures; 
(ii) a $T$-independent isosbestic line with linear energy dependence; 
(iii) SW transfer from below to above this line; 
and (iv) similar absolute values for $\sigma_1(\omega)$ in theory and experiment. 

While the match between our first-principles calculations and the experiment is very good, the question remains whether we can (i) isolate the contribution of the DNL and (ii) explain such an unusual isosbestic line.
To answer these two questions, we constrain the $k$-integration range in the current-current response function calculation to a cylinder in the IBZ (shaded volume in Fig.~\ref{fig2}(a)). 
This cylinder selects intra and interband transitions involving states that mainly belong to the Dirac cone centered at $k_\textrm{Dirac}$.
When the radius R$_\textrm{sel}$ of the cylinder is set to 1/4 in crystal units (Figs.~\ref{fig2}(c--d)),  $\sigma_1 (\omega)$ becomes flat between 0.2 and 0.4 eV, and its second peak at \mev{650} disappears. 
By further shrinking R$_\textrm{sel}$ to 0.2, to better isolate the DNLs around $k_\textrm{Dirac}$, the peak at \mev{550} is suppressed as well, while the plateau in the 200--\mev{400} range remains (Fig.~\ref{fig2}(d)).
The flat and $T$-independent response of $\sigma_1(\omega)$ (Fig.~\ref{fig2}(c)) is characteristic of a 2D conical dispersion~\cite{Gusynin2006, Kuzmenko2008, Mak2008, Nair2008}. 
This is a strong first-principles evidence that the isosbestic character of $\sigma_1(\omega)$ from 0.2 to 0.4 eV is caused by the electronic states involved in the dispersive nodal line.
The same Dirac states reveal their conical in-plane dispersion in our magneto-optic measurements.

However, this does not yet explain the actual slope seen in both experiment and full, unrestricted $\sigma_1(\omega)$ calculations over that energy window. 
To solve this mismatch, we also analyze the $\sigma_1(\omega)$ contributions coming from transitions around $\Gamma$ and  X, excluded in the previous $k$-selection in Fig.~\ref{fig2}(c).
We plot the results of $k$-selected transitions delimited by cylinders of radius $R_\textrm{sel}=0.2$ centered at $\Gamma$ and X in Fig.~\ref{fig2}(d).
The calculated $\sigma_1$ with $k$-points from $\Gamma$ perfectly covers the 200--\mev{450} frequency range, with a slope very close to full $\sigma_1$. 
This linear $\sigma_1(\omega)$ contribution can be explained in the band structure by the presence of a second conical shape centered at $\Gamma$, which is connected to the bands of the DNL at the \GM\ line mid-point, Fig.~\ref{fig3}(a).
The contribution of the $k$-points around X starts at \mev{300}. 
By dissecting the IBZ into distinct patches, we prove that the isosbestic line is made out of a flat and strictly $T$-independent part coming from the response of the DNL at the \GM\ line mid-point; and a higher-energy linear contribution from conical states near $\Gamma$.
The former contribution is typical of the 2D and 3D conical model, whose optical response becomes temperature independent at an energy far enough from the optical gap~\cite{Falkovsky2007,Falkovsky2007B,Stauber2008, Carbotte_temp_2016}
The isosbestic line is then the optical signature of purely conical excitations. 
In \bns\, the isosbestic energy range becomes clearly visible because of the quasi-2D nature of the material. 
This leads to open Dirac lines developing along $k_z$ such that, in a first approximation, one can see the system as made of infinite copies (in $k_z$) of a 2D Dirac system.

%
%
We will now focus on the low-temperature optical conductivity at energies below the isosbestic line.
An ideal 2D Dirac cone with non-$k_z$-dispersing line nodes would have a constant $\sigma_1(\omega)$ \cite{Kuzmenko2008}, and would not explain the dip and the steep slope of $\sigma_1(\omega)$ below the nearly flat plateau. 
These features come from the $k_z$-dispersing nodal lines in \bns\, as shown in Fig.~\ref{fig3}(a). 
The plotted gapped Dirac nodes projected in the \GM\ direction as a function of k$_z$ are obtained from the HSE band structure in Fig.~\ref{fig2}(b). 
This can be modeled by the energy dispersion \cite{Shao2019}:
\begin{equation}
\epsilon_{\pm} =\pm\sqrt{\Delta^{2}+v_{1}^{2}k_{1}^{2}+v_{2}^{2}k_{2}^{2}}+v_z k_z ,
\label{eq2}
\end{equation}
where $k_{1}$ and $k_{2}$ are momenta perpendicular to the nodal line, $\Delta$ is the SOC-induced gap, and $v_i$ the in-plane asymptotic velocities that describe the slope of the Dirac cone.
The HSE band structure gives the asymptotic velocities $v_1=2.61$ \ev{}\AA\ and  $v_2=0.94$ \ev{}\AA.
Their average is 1.57~\ev{}\AA, fairly close to the experimental $v=1.02$~\ev{}\AA.

$\sigma_1(\omega)$ associated with the dispersion in Eq.~\ref{eq2} is \cite{Shao2019}:
\begin{equation}
\sigma_{\it NL}^{i}(\omega) =\dfrac{N}{8}\dfrac{e^{2}}{h}k_{0}(\omega)\dfrac{v_{i}^{2}}{v_{1}v_{2}}\left(1+\dfrac{4\Delta^{2}}{\omega^{2}}\right) \Theta\left(\omega - 
2\Delta
\right) , 
\label{eq3}
\end{equation}
where $N$ is the number of nodal line (including the spin multiplicity) in the IBZ. The electric field is polarized along the $i$ direction.
The length $k_0(\omega)$ is the chunk of the nodal line accessible at a given $\omega$, see the supplementary for its photon energy dependence \cite{SM}. 
We take Eq.~\ref{eq3} with the electric field averaged in plane, and plug in the magneto-optical parameters. 
We also fit the energy shift of the nodal line along $k_z$, so as to match the experimental onset of interband excitations.
All of this gives us a $\sigma_{\it NL}(\omega)$ dependence, shown in Fig.~\ref{fig3}(b) as a dashed black line. 
This curve follows the experimental data well up to \mev{80}, but fails to reproduce the steeper isosbestic slope at higher energies. 
In contrast, $\sigma_{\it NL}(\omega)$ behaves much more like the \emph{ab initio} calculations restricted around $k_\textrm{Dirac}$. 
This confirms that the dip, the slope and the isosbestic flat region of $\sigma_1(\omega)$ come from DNLs dispersing along $k_z$.

Finally, we want to explain the experimental peak in $\sigma_1$ at \mev{350}, captured by the \emph{ab initio} peak at \mev{550}.
We developed a simplified 2D tight-binding model on a square lattice \cite{SM, berthod2013, ando2002,min2009,mukherjee2017}, in line with the quasi-2D nature of \bns.
The electronic structure of this model is plotted in Fig.~\ref{fig3}(c). 
The fourfold Dirac cones, like in \bns, are connected by arches whose topmost part leads to VHSs in the DOS, and to the peak found in the computed $\sigma_1(\omega)$ at $\approx$ 1.1 eV, Fig.~\ref{fig3}(d). 
This toy model shows that the higher energy peaks of $\sigma_1(\omega)$, looming over the flatter isosbestic regions are caused by the connecting Dirac cones within the IBZ. 
The cones' straight walls eventually bend to merge with one another. 
These VHSs are thus universal in a system of Dirac cone networks~\cite{Mak2011,Mak2014,santos2020,Ebad2019, Martino2019, Mardele2020}. 

%
%
In conclusion, we measured the optical conductivity signatures of the open dispersive nodal lines dominating the low-energy spectrum of \bns\, and a VHS at a higher energy, with an isosbestic line connecting both regions. 
The magneto-optical response indicates a slightly gapped in-plane linear dispersion.
Using first-principles calculations we fully reproduce the experimental optical spectra and we reverse engineer them. We show that \bns\ is an exemplary nodal line Dirac semimetal, whose band structure is simpler than the more investigated counterparts, and with strong 2D features, despite being a 3D layered material. 
This simplicity allowed us to uncover the optical signatures of conical dispersions, and the impact of their $k_z$ and temperature dependence. 
The observed isosbestic line and the VHS are universal fingerprints of untainted quasi-2D Dirac excitations in the low-energy range. 
The VHS also marks the conical model breakdown when the Dirac cones connect in reciprocal space.
This peculiar optical response is thus an experimental proxy which can unequivocally identify new beyond-graphene materials.


M.C. acknowledges 
GENCI allocations for computer resources under the
project number 0906493.
L.d.M is supported by the European Commission through the ERC-CoG2016, StrongCoPhy4Energy, GA No 724177.
A.~A. acknowledges funding from the  Swiss National Science Foundation through project PP00P2\_170544.
This work was supported by the ANR DIRAC3D project (ANR-17-CE30-0023).


%

\end{document}